\documentclass[traditabstract]{aant} % for the abstract without structuration  % (traditional abstract)
\usepackage{graphicx}
\usepackage{txfonts}
\usepackage{color}
\usepackage{wasysym}
\usepackage{natbib}
\begin{document}

    \title{\vspace*{-2.0cm} Seeding grain nucleation and dust growth}
    \subtitle{Ionisation, epoxidation and charge disproportionation effects}

    \author{A.P. Jones}
     \institute{Universit\'e Paris-Saclay, CNRS,  Institut d'Astrophysique Spatiale, 91405, Orsay, France.\\
               \email{anthony.jones@universite-paris-saclay.fr}
%         \and
%           University \\
%             \email{ }
%             \thanks{}
              }

    %\date{Received ??? 2025; accepted ??? 2025}

   \abstract
% context
{Dust nucleation and growth is not yet completely understood and this is particularly the case in the partially or weakly ionised media typical of the nov\ae,  Wolf-Rayet, and Luminous Blue Variable systems where dust formation occurs on a relatively short timescale. Further, the onset of dust growth in the outer regions of molecular clouds, presumably aided by gas phase accretion and nanoparticle sticking, appears to be efficient but hard to explain.}
%aims
{This work studies the likely dust seeding processes arising from alkali metal and alkaline earth ionisation, epoxidation (epoxide bond formation via oxygen atom insertion into C$=$C bonds), and grain charge disproportionation (the existence around the uncharged state of oxidised cationic and reduced anionic states) at (sub-)nanometre size scales.} 
% methods
{The chemical, physical, and photon-initiated processes leading to dust seeding are explored within the framework of the size-dependent physical, optical, and photoelectric properties of the THEMIS carbonaceous nanoparticles. The critical grain charge states at (sub-)nanometre size scales are derived as a function of the interstellar and circumstellar physical conditions.}
% results
{Photo-initiated low-energy ionisation, epoxide reactions, and disproportionation-driven electrostatic effects could play key roles in seeding dust nucleation and growth. The size-dependent seed cluster and nanograin charge distribution is shown to encompass both positive and negative charges where the ionisation is driven by low ionisation metals or by weak attenuation.} 
% conclusions
{Cluster seeding via ionisation and epoxidation could help to explain the co-spatial and contemporaneous nucleation and growth of both carbon-rich and oxygen-rich dust in the same regions. This may be enhanced by electrostatic effects, driven by charge disproportionation, between negatively-charged, nucleation-seeding, polyatomic clusters and positively-charged ions or larger (nano)particles. Such processes could occur in the dust-forming regions in nov\ae,  Wolf-Rayet, and Luminous Blue Variable systems and electrostatic effects may also aid the accretion of nanoparticles in the outer regions of molecular clouds.}
   %\keywords{ISM: abundances -- ISM: dust,extinction               }
   \keywords{\vspace*{0.2cm}}

    \maketitle
    \nolinenumbers
%
%________________________________________________________________

%-------------------------------------------------------------------------------------------------------------------------------- 
\section{Introduction}
%-------------------------------------------------------------------------------------------------------------------------------- 

The study of the mechanisms of dust nucleation and small molecule formation in circumstellar shells has a long history \cite[e.g.][]{1968ApJ...153..451D,1984A&A...132..163G,1984A&A...133..320G,1989ApJ...341..372F} primarily using thermodynamic and kinetic approaches. Since these early works numerous studies have followed and developed these principles \cite[e.g.][]{2012MNRAS.420.3344G,2013ApJ...776..107S,2015A&A...575A..95S,2016A&A...591A..17G}. More recent work, using density functional theory techniques, focussed on the thermodynamic stabilities of the generic metal oxide clusters MO$_n$ (where M = Al, Mg, Si, Ca, and/or Ti) with stoichiometries corresponding to $n = 1.33$ or 1.5 have been undertaken \cite[e.g. see][and references therein]{10.3389/fspas.2025.1632593}. Nevertheless, our understanding of the details of dust nucleation and the onset of dust growth is still incomplete. 

This work focusses on the photo-initiated atomic-scale processes that may be responsible for the initial seeding of dust-forming nuclei or clusters, hereafter called seed clusters. Photon-driven processes, the hub around which most interstellar medium (ISM) studies revolve, are likely at the very heart of seed cluster formation and evolution. This work tackles the challenge of understanding the rapid formation of dust in nov\ae,  Wolf-Rayet (WR) and Luminous Blue Variable (LBV) systems and attempts to explain how they can  co-spatially and contemporaneously form both oxygen-rich and carbo-rich dusts. The photo-sensitive processes of epoxidation and charge disproportionation, coupled with the presence of low ionisation metal ions, would appear to be pivotal for the seeding of dust in these environments.

Epoxidation is the process that leads to the formation of a triangular epoxide ring (a three atom C$-$O$-$C cyclic structure, ${\rm C}-\hspace{-0.3cm}^{\rm O}\,{\rm C}$) through the addition of electronegative (electron-grabbing) oxygen atoms to electron-rich carbon-carbon double bonds, C$=$C. Epoxide ring structures are stable but will open up upon blue-violet photon irradiation to form new bonds. 
%, the process that is at the heart of epoxy adhesives. 

Disproportionation is a term used in chemistry for a redox reaction in which a substance of a given oxidation state transforms into two others of correspondingly higher and lower oxidation states. Using the chemical definition of oxidation as the loss of electrons and reduction as the gain of electrons, the term disproportionation is used here in relation to electrically neutral particles. With reference to an initially uncharged particle, oxidation is then the loss of an electron, via photoelectron emission, to form a cationic grain (G$^0 + {\rm h}\nu \rightarrow$ G$^+ + e^-$) and reduction is the gain of electrons to form anionic species (G$^0 + e^- \rightarrow$ G$^-$). With this definition disproportionation is used in relation to oppositely charged grains within a given environment. 

This work is concerned with the nano-physics of dust seeding and nucleation at the very onset of dust formation via the interaction of atoms, ions and nano-clusters to form somewhat larger species. The key processes are akin to accretion, a process generally involving particles with large mass differences, that is at the heart of heterogeneous carbonaceous mantle formation in the dust modelling framework THEMIS \citep[The Heterogeneous dust Evolution Model for Interstellar Solids,][]{2013A&A...558A..62J,2014A&A...565L...9K,2017A&A...602A..46J,2024A&A...684A..34Y}.\footnote{See  http://www.ias.u-psud.fr/themis/} 

Nov\ae,  WR and LBV systems are interesting cases because they can be fast and efficient dust producers but the details of how they can form dust so efficiently are still not satisfactorily explained. \cite{Shore:2004hj} developed an elementary model for grain growth in nov\ae\ arguing that nucleation is a kinetic process where induced dipole interactions play a critical role and lead to runaway growth in a partially ionised medium. 

One of the most recent surveys of dust formation in nov\ae\ \citep{Shore:2018bi} points out that dust is generally not present before the beginning of the steep photometric decline, but grows rapidly with the onset of transparency in the ultraviolet (UV).  Further, after the maximum in the opacity the dust optical depth decreases due to expansion-driven dilution alone. This same study underlines the observational inference that nova-formed grains are large (radii $\geqslant0.1\mu$m) and,  following the explosion, the newly-formed dust survives exposure to the intense X-ray irradiation with little change in its properties long after the photometric minimum is passed. Thus, nova dust survives and mixes with the interstellar gas at late times. 
%but does, nevertheless, appear to remain a minor constituent of interstellar dust.  

Given that about one third of classical novae produce dust it has been tantalisingly difficult to unambiguously identify any presolar stardust grains from such sources \cite[e.g.][]{2018ApJ...855...76I}. However, some of the isotopic signatures measured in a handful of presolar grains do seem to qualitatively reproduce the properties expected from the explosive burning of hydrogen in classical novae \citep{2001ApJ...551.1065A}. A grain of unequivocal of CO nova origin was identified by \cite{2019NatAs...3..626H}; albeit one that poses some fundamental questions about dust formation and the C/O ratio as a predictor of the dust composition. Despite the fact that the overall contribution of nova and WR dust to the interstellar dust budget is certainly $< 1$\%
\cite[e.g.][]{2004ARA&A..42...39C,2014mcp..book..181Z}, the rapid formation of dust in these environments imposes critical constraints on dust formation in these and, in general, on all circumstellar dust forming environments. 

Possibly related to the question of rapid dust nucleation in evolving circumstellar regions is the issue of the onset of dust growth in the interstellar medium (ISM) or cold neutral medium (CNM) through the disappearance of nanoparticles in denser regions. This is evidenced by a $80-100$\% decrease in the 60$\mu$m emission coming from these stochastically heated nano-grains, with radii of the order of $1-10$nm, in the transition from diffuse to dense molecular clouds \citep{1991ApJ...372..185L,1996A&A...309..245A,1994ApJ...423L..59A,1998A&A...333..709L,2003A&A...399.1073K,2003A&A...398..551S,2012A&A...548A..61K}. It is likely that some as yet unidentified process may be accelerating the loss of nanoparticles in these regions, almost certainly through their accretion onto larger grains, but such a process has yet to be elucidated. 

This paper is organised as follows: 
Section \ref{sect_nucleation} sets the scene for dust seeding in nov\ae\ and elsewhere, 
Section \ref{sect_charge} presents the calculation of the seed cluster and nanoparticle charge states, 
Section \ref{sect_seeding} proposes a scenario for seedeing dust nucleation and growth, and
Section \ref{sect_conclusions} discusses the results and concludes.

%-------------------------------------------------------------------------------------------------------------------------------- 
\section{Dust seeding in nov\ae\ and elsewhere}
\label{sect_nucleation}
%-------------------------------------------------------------------------------------------------------------------------------- 

The seeding and nucleation of dust particles around evolving stars is a perennial problem that is particularly difficult in the case of the rapid dust formation around nov\ae\ \citep[e.g.][]{Shore:2004hj} and also WR stars \citep[e.g.,][]{2023ApJ...951...89L}. These environments present a particular challenge for dust nucleation modelling because of the rapidly evolving radiation field and density, post explosion or eruption, and our incomplete understanding of dust nucleation and growth under such conditions. In the case of nov\ae\ it is also something of a puzzle as to how both oxygen-rich and carbon-rich dusts can form at the same time and in the same place. Laboratory evidence from pre-solar grains, in the form of a graphitic grain originating from a low-mass carbon- and oxygen-rich nova that encompasses both silicate and oxide nanoparticles, attests to the co-formation of these dusts amongst nova ejecta \citep{2019NatAs...3..626H}. As underlined in the work of \cite{Shore:2004hj} the ionisation state of the gas is likely to be critical. It is therefore important to understand the interactions between, and the respective roles of, the local stellar radiation field (its mean and maximum photon energies), the chemical composition and space density of the gas, the ionisation potentials (IPs) of the constituent atoms, and the charge distribution across the nucleating and growing grains in the plasma. 

The usual dust forming elements of C, Si, Mg, Fe, and O have rather high ionisation potentials of 11.26, 8.15, 7.65, 7.90, and 13.61eV, respectively. In contrast, the alkali metal and alkaline earth ions K$^+$, Na$^+$, Li$^+$ and Ca$^+$, with production potentials (from the neutral parent atoms) of 4.34, 5.14, 5.39, and 6.11eV, respectively, can be ionised in low ionisation regions where the usual dust forming atoms are neutral. Low ionisation in this sense refers to the relatively low mean energy of the ionising photons (i.e. $\langle E_{\rm h \nu}\rangle \ll 13.6$eV) rather than to the fractional ionisation of the gas which may be extremely low at the nucleation epoch, possibly as low as $\sim 10^{-5}$ if it is solely governed by the complete ionisation of the low abundance and low ionisation potential metals mentioned above.  It is therefore possible that K$^+$, Na$^+$, Li$^+$, and Ca$^+$ ions could play a role in regions dominated by low mean photon energy, $\langle E_{\rm h \nu}\rangle \lesssim 6$eV, radiation fields. 
This may indeed be the case at the onset of the dust forming epochs around nov\ae\ and WR stars when the radiation field is hardening as the light curve declines but the mean ionising photon energy is still low ($\lesssim 6$eV), conditions that seem to be associated with the start of dust formation in these systems \citep[e.g.][]{Shore:2018bi}. 

In summary, the presence of K$^+$, Na$^+$, Li$^+$, and Ca$^+$ ions provide not only the critical dust-nucleating centres but are also the source of the electrons in the gas. The dominating contribution of these ions is entirely dependent upon a low mean photon energy, $\langle E_{\rm h \nu}\rangle \lesssim 6$eV, radiation field in which the typical dust-forming elements (C, Si, Mg, Fe, and O) remain neutral. Such an elemental ioisation stratification would appear to be critical for seeding dust nucleation and growth.

%-------------------------------------------------------------------------------------------------------------------------------- 
\subsection{Cluster-metal ion interactions}
\label{sect_nucleation_a}
%-------------------------------------------------------------------------------------------------------------------------------- 

The progressive association reactions of alkali metal and alkaline earth ions, here generically represented by $X^+$,  with neutral carbon atoms would be expected to proceed along the lines 
\begin{equation}
X^+ + {\rm C} \rightarrow X{\rm C}^+ \ \ \ \ \ \ \ \  X{\rm C}^+ + {\rm C} \rightarrow X{\rm C}_2^+   \ \ \ \ \ \ ...  \ \ \ \ \ \rightarrow X{\rm C}_n^+. 
\label{eq_scheme_1a}
\end{equation}
As pointed out by \cite{Shore:2004hj} induced dipole interactions in atom-grain collisions are likely to play an important role in enhancing collision rates in partially ionised media. This same effect will be even more enhanced in ion-neutral reactions where the neutral specie is a small ($a < 0.3$nm), polarisable dust nucleating ${\rm C}_n$ cluster (where $n$, the number of carbon atoms is $\lesssim 20$). This is directly analogous to the ion-neutral gas phase reactions that are at the core of interstellar chemistry and which drive molecule formation in the ISM. 

In dust nucleating interactions there will almost certainly be preferred values of $n$ associated with more stable carbon cluster configurations, for example, but certainly not exclusively, those for small aromatics (e.g. $n = 6$, 10, 14, \ldots) or fullerenes (e.g. $n = 20$, 26, 28, \ldots).  

\begin{figure}
 \includegraphics[width=9.5cm]{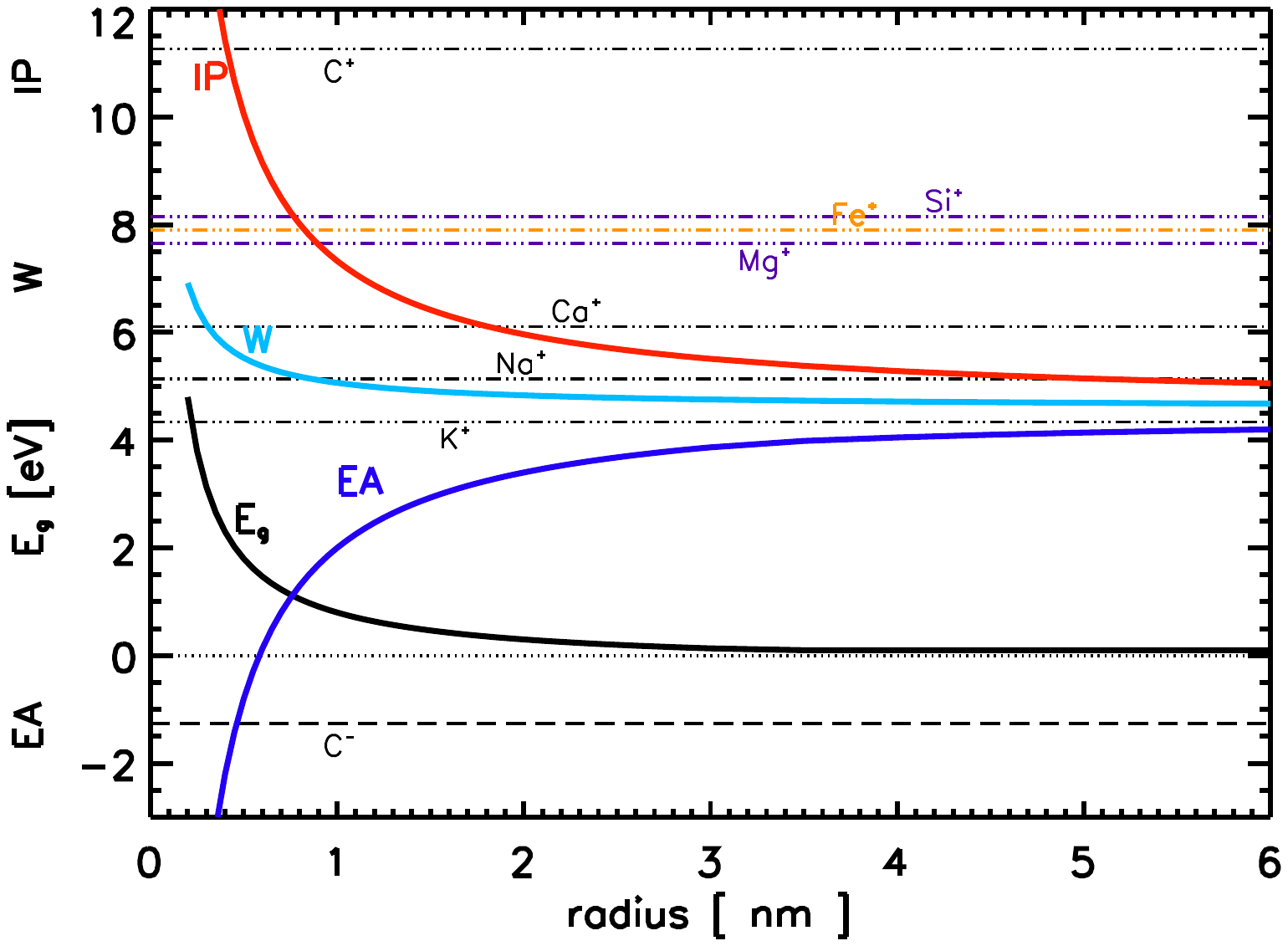}
 \caption{Size-dependent nanoparticle IP (red line, Eq. \ref{eq_IP}), EA (blue line, Eq. \ref{eq_EA}), work function ($W$, cobalt line, Eq. \ref{eq_Wa}) and band gap  ($E_{\rm g}$, black line, Eq. \ref{eq_Eg}). The atomic IPs and the carbon atom EA (dashed and dotted horizontal lines) are also shown. 
 %The production potentials for  Si$^+$, Fe$^+$, Mg$^+$, Ca$^+$, Na$^+$, and K$^+$ are indicated by the labelled horizontal dashed triple-dotted lines. 
 All quantities are in eV. } 
\label{fig_params}
\end{figure}

Recombination of the cationic clusters via electron collisions will likely re-form the catalysing ion $X^+$, where it has a lower ionisation potential than the C$_n$ cluster (as is the case for $a \leqslant 2$nm, see Fig. \ref{fig_params}), that is 
\begin{equation}
X{\rm C}_n^+ + e^- \rightarrow   {\rm C}_n^- + X^+. 
\label{eq_scheme_1b}
\end{equation}
The formation of small anionic clusters, C$_n^-$, will be aided by their electron affinity (EA $< 0$\,eV  for $a \leqslant 0.6$nm, Fig. \ref{fig_params}). 

In contrast a neutral carbon cluster C$_n$ can undergo charging via electron and ion collisions and photon absorption leading to photoelectric emission, that is 
\begin{equation}
{\rm C}_n + e^- \rightarrow {\rm C}_n^-,  \ \ \   {\rm C}_n + X^+ \rightarrow X{\rm C}_n^+, \ \ \ {\rm and}  \ \ \ 
{\rm C}_n + {\rm h}\nu \rightarrow {\rm C}_n^+ + e^-,  
\label{eq_scheme_1c}
\end{equation}
respectively. The middle reaction above can only occur if the ion has a non-zero probability of sticking to the cluster and for small clusters this will be aided by the same microphysical processes, such as radiative association, that drive ion-molecule reactions. Charge exchange cannot occur with the smallest dust-nucleating clusters in low excitation radiation regions because the most probable partner ions, K$^+$, Na$^+$, Li$^+$, and Ca$^+$, all have ionisation potentials less than that for sub-nanometre radius carbon clusters (IP $\gtrsim 7$eV for $a \lesssim 1$nm, see Fig. \ref{fig_params}). Further, the smallest clusters have relatively high electron affinities and will therefore have a strong tendency to be negatively charged. 

A dust-nucleating seed cluster may be defined as any stable small, neutral or charged, molecule, radical, ion or radical ion, typically containing less than several tens of atoms, that acts as a critical centre for dust growth. It could be any species such as: C$_n^{(-/0/+)}$ ($n \lesssim 30$), MC$_m^{(-/0/+)}$ or MO$_m^{(-/0/+)}$ where M is a metal atom (Li, Na, K, Ca, Ti, Al, Mg, Si, or Fe) where $m$  ranges $0.5-2$. 

It seems likely that at some critical size  the electron recombination of a cluster, which may or may not include a metal atom, $(X){\rm C}_n^+$, will simply result in charge neutralisation, which can then be followed by electron attachment, that is 
\begin{equation}
X_m{\rm C}_n^+ + e^-   \rightarrow  \ X_m{\rm C}_n \ \ \ \ \ \  X_m{\rm C}_n + e^- \rightarrow  \ X_m{\rm C}_n^- \\\\ m \geq 0. 
\label{eq_scheme_1bc}
\end{equation}
Anionic clusters can accrete metal ions in Coulomb focussed interactions, which could then react with gas phase oxygen atoms (see the following sub-section) leading to the formation of separate carbon-rich and oxygen-rich dust precursors in the same regions 
\begin{equation}
X_m{\rm C}_n^- + X^+\rightarrow    X_{m+1}{\rm C}_n \ \ \ \ \ \ \ \  X_m{\rm C}_n^{(-)} + {\rm O} \rightarrow {\rm C}_n + X_m{\rm O}^{(-)}. 
\label{eq_scheme_1bd}
\end{equation}
The left reaction above will be Coulomb enhanced by factors of roughly 170, 70, 30, and 20 for Ca$^+$ ions incident on C$_n^-$ anions of radii $\simeq 0.2$, 0.5, 1, and 2nm, respectively, at a gas temperature of 1000K. The right hand reaction will be aided by induced dipole interactions \citep[e.g.][]{Shore:2004hj} or where the cluster is anionic ($X_m{\rm C}_n^-$) driven by an ion-neutral interaction. 
%by the equivalent of an ion-neutral reaction as is standard in interstellar chemistry models. 

%-------------------------------------------------------------------------------------------------------------------------------- 
\subsection{Cluster-atomic oxygen interactions}
\label{sect_nucleation_b}
%-------------------------------------------------------------------------------------------------------------------------------- 

Given that ethyne, HC$\equiv$CH, and ethene, H$_2$C=CH$_2$, are common molecules in dust forming regions \citep[e.g., around IRC+10216,][]{2006PNAS..10312274Z}, which may also contain oxygen atoms in neutral form, as also found in nov\ae, it would appear that to date not all of the possible reaction pathways between these hydrocarbon species and oxygen atoms have been considered. In particular, electronegative oxygen atoms are known to readily react with C$\equiv$C and C=C bonds. In the case of ethene this occurs by O atom insertion directly into the C=C bond to form an epoxide  that has a triangular structure with two C atoms and an O atom at the apices, here represented by C$-$\hspace{-0.2cm}$^{\rm O}$C, and two H atoms bonded to each C atom, that is H$_2$C$-$\hspace{-0.2cm}$^{\rm O}$CH$_2$  known as ethylene oxide or oxirane. This molecule is produced during the incomplete combustion of ethene and is the only alkene molecule to form an epoxide in this way. Epoxide species are very reactive and readily decompose upon blue-violet photolysis by ring-opening to form new bonds; the underlying mechanism for bond formation in epoxy glues. The richness of the products of the atomic O reaction with C=C bonds leading to epoxides and their subsequent photolysis, was explored in some detail within the astrophysical context by \cite{2016RSOS....360221J,2016RSOS....360224J}. 

In the following it is assumed that, at the temperatures and physical conditions of the ISM and circumstellar media, the reaction between atomic oxygen and $>$C=C$<$ bonds is efficient, that there are abundant metal cations ($X^+$) present with lower ionisation potentials ($4-6$\,eV, see above) than carbon clusters (IP $\gtrsim 6$eV for radii $\lesssim 2$nm), and that the electron affinity of the carbon clusters is less than the cation production potential. With these explicit assumptions, the following generic reaction cycle is proposed for the formation of epoxide bonds, C$-$\hspace{-0.2cm}$^{\rm O}$C, and cations bound to C=C bonds, that is C=C$\cdot\cdot(\cdot) X^+$ where fewer dots indicate stronger ion-induced dipole interactions:
\[
{\rm C}={\rm C}  \ \ \rightarrow^{\hspace{-0.3cm} X^+}  \ \ {\rm C = C}\cdot\cdot X^+ \ \ \rightarrow^{\hspace{-0.3cm} \rm O}  \ \ \ {\rm C-^{\hspace{-0.2cm} \rm O} C}\cdot\cdot X^+
\]
\[
{\rm C}={\rm C}  \ \ \rightarrow^{\hspace{-0.3cm} \rm O}  \ \ \ {\rm C-^{\hspace{-0.2cm} \rm O} C} \hspace{1.0cm}  \rightarrow^{\hspace{-0.3cm} X^+}  \ \  {\rm C-^{\hspace{-0.2cm} \rm O} C}\cdot\cdot X^+
\]
\[
 {\rm C-^{\hspace{-0.2cm} \rm O} C}\cdot\cdot X^+ \ \ \ \ \rightarrow^{\hspace{-0.3cm} \rm h \nu}  \ \ \ {\rm C}={\rm C}^+ \ + \ X{\rm O} 
 \] 
\begin{equation}
{\rm C}={\rm C}^+ \ + e^- \ \rightarrow \ \ \ {\rm C}={\rm C}.
\label{eq_scheme_2a}
\end{equation} 
For anionic clusters, C=C$^-$,  the above reactions become   
\[
{\rm C}={\rm C}^-  \ \ \rightarrow^{\hspace{-0.3cm} X^+}  \ \ {\rm C = C}\cdot\cdot\cdot X \ \ \rightarrow^{\hspace{-0.3cm} \rm O}  \ \ \ {\rm C-^{\hspace{-0.2cm} \rm O} C}\cdot\cdot\cdot X
\]
\[
{\rm C}={\rm C}^-  \ \ \rightarrow^{\hspace{-0.3cm} \rm O}  \ \ \ {\rm C-^{\hspace{-0.2cm} \rm O} C}^- \hspace{0.9cm}   \rightarrow^{\hspace{-0.3cm} X^+}  \ \  {\rm C-^{\hspace{-0.2cm} \rm O} C}\cdot\cdot\cdot X
\]
\[
 {\rm C-^{\hspace{-0.2cm} \rm O} C}\cdot\cdot\cdot X \ \ \ \rightarrow^{\hspace{-0.3cm} \rm h \nu}  \ \ \ {\rm C}={\rm C} \ + \ X{\rm O} 
 \] 
\begin{equation}
{\rm C}={\rm C} \ + e^- \ \ \, \rightarrow \ \ \ {\rm C}={\rm C}^{-}.
\label{eq_scheme_2b}
\end{equation} 
Where, for simplicity in the above reactions, the bonds to hydrogen atoms and/or side groups, and the atoms and groups themselves, are not shown. In reality the C=C bond in the above schemes will be a sub-component of a larger carbon seed cluster, which will stabilise the bond against dissociative recombination, the last step in Eq. (\ref{eq_scheme_2a}). In these cases the C=C bonds within the carbon clusters are regenerated and therefore act as a catalysts for metal oxide, $X$O, formation, where $X$ is most likely to be a K, Na, Li, or Ca ions but may also be Mg, Fe, Si, Al, or Ti atoms. It is of note that the photo-initiated opening of an epoxide (oxirane) ring typically requires blue-violet photons ($\lambda \sim 430$nm) of energies ($E_{\rm h \nu} \sim 2.9$eV)\footnote{The exact triggering photon energy is therefore not that critical because it is low compared to the metal ionisation energies.} significantly less than the critical ionisation potentials of K, Na, Li, and Ca atoms ($E_{\rm h \nu} \sim 4-6$eV).

Although the two sets of reactions, Eqns. (\ref{eq_scheme_1a}) to (\ref{eq_scheme_1bd}) and (\ref{eq_scheme_2a}) and (\ref{eq_scheme_2b}), are presented separately they will occur at the same time and place for the same cluster, ion and atomic species.

% FIGURE *********************************************************
%  *********************************************************

% TABLE *********************************************************
\begin{table*}
\caption{Photon energies and fluxes, gas densities and temperatures taken to be typical of nov\ae, WR stars etc. The colours indicated in the column headers refer to Fig. \ref{fig_charge}. The CNM cloud disproportionation radii are $\simeq 1.2$, 1.5, 2.2, 7, 200, and 700nm for $A_{\rm V} =$ 0, 0.25, 0.5, 1, and 2, respectively.}
\begin{center}
\begin{tabular}{lccccc}
      &         &        &      &  &  \\[-0.35cm]
\hline
\hline
      &         &        &      &  &  \\[-0.35cm]
  & \multicolumn{3}{c}{nov\ae\ \& WR systems}    & \multicolumn{2}{c}{Interstellar media} \\[0.05cm] 
  case $\rightarrow$     &  A  (purple)                     &    B  (blue)                  &   C  (cobalt)                & CNM (red)& cloud (orange) \\[0.05cm]
\hline
                                                    &                                    &                                     &                                   &                           &               \\[-0.30cm]
mean photon energy  ( eV )             &             5.8                  &              5.9                   &          6.0                    &        7.0               &         7.0               \\
photon flux    ( photons/cm$^2$/s )  &  $7.5 \times 10^{11}$  &   $8.0 \times 10^{11}$   & $8.5 \times 10^{11}$  &  $3 \times 10^7$  &  $1 \times 10^7$  \\
gas density ( cm$^{-3}$ )                &   $6 \times 10^{7}$     &    $4 \times 10^{7}$       &  $2 \times 10^{7}$     &       100                &       2,000             \\
gas temperature ( K )                       &       1,000                    &          1,000                   &        1,000                   &      100               &          20               \\ 
disproportionation radius [nm]           &        150                 &          20                 &        3                 &     ---          &       see caption             \\
\hline
\hline
     &      &         &       &  &  \\[-0.25cm]
\end{tabular}
\end{center}
\label{table_conds}
\end{table*}
% *********************************************************

% TABLE *********************************************************
\begin{table}
\caption{Collision enhancement factors, $F_{\rm C}$ for the charge states shown in Fig. \ref{fig_charge}), for the physical conditions in Table \ref{table_conds}.}
\begin{center}
\begin{tabular}{cccccc}
 &   &   &   &  &    \\[-0.35cm]
\hline
\hline
 &   &   &   &  &    \\[-0.35cm]
Nov\ae\ \& WR       &     $F_{\rm C}$      &      $a_1$ [nm]    &     $Z_1$   &    $a_2$ [nm]   &     $Z_2$   \\[0.05cm] 
\hline
 &   &   &   &  &   \\[-0.30cm]
cobalt   &      5.0     &    0.3   &      -1      &      3       &       +1      \\
cobalt   &      3.5     &    0.3   &      -1      &      5       &       +1      \\
cobalt   &      3.2     &     1     &      -1      &      40      &       +7     \\
blue      &      1.3     &    1     &     -1      &       40      &       +1      \\
purple   &      1.3     &    1     &     -1      &       40      &       +1       \\
\hline
 &   &   &   &  &  \\[-0.35cm]
\multicolumn{6}{l}{Cloud A$_{\rm V}$  (orange)}  \\[0.05cm] 
\hline
 &   &   &   &  &   \\[-0.35cm]
   0           &       2.7     &     0.4     &     -1      &       150      &       +2        \\
   0           &       2.7     &      1     &     -1      &       150      &       +2        \\
$0-1$       &     25.3     &     0.4     &     -1      &        5        &       +1        \\
$0.25-1$   &       1.9     &     0.4    &     -1       &      150       &       +1       \\
$0.25-1$   &       1.9    &       1     &     -1       &      150       &       +1        \\
$2-3$       &      0.1     &     0.4     &     -1      &      150       &        -1       \\
\hline
\hline
 &   &   &   &  &   \\[-0.25cm]
\end{tabular}
\end{center}
\label{table_F_Coulomb}
\end{table}
% *********************************************************

%-------------------------------------------------------------------------------------------------------------------------------- 
\section{Seed cluster and nanoparticle properties}
\label{sect_charge}
%--------------------------------------------------------------------------------------------------------------------------------    

In this study we adopt the optical and physical properties of the (hydro)carbonaceous, a-C(:H), nanoparticles of the THEMIS interstellar dust model \citep{2013A&A...558A..62J,2017A&A...602A..46J,2014A&A...565L...9K,2024A&A...684A..34Y}. The optical properties are, at nanometre size scales, principally determined by the electrons associated with the largest contiguous aromatic systems within these amorphous semi-conducting materials; that is by their sp$^2$ $\pi$ electrons. The size-dependent particle band gap, $E_{\rm g}$, is given by  \cite{2012A&A...542A..98J} 
\begin{equation} 
E_{\rm g} {\rm [eV] = max} \left\{ -0.18, \ 0.2\left[ \left(\frac{5\,{\rm nm}}{a} \right)-1 \right] \right\}.  
\label{eq_Eg} 
\end{equation}
Thus, $E_{\rm g}$ increases with decreasing particle size, coherent with a transition to the molecular nature of sub-nanometre particles, and exhibits a minimum of $-0.18$\,eV for large, aromatic-rich particles \citep[e.g.,][]{1984JAP....55..764S}, as shown in Fig.~2 of \cite{2012A&A...542A..98J}.

For a-C(:H) (nano)particles the thresholds for the photo-electric effect and IP are set by the least binding states of the electrons, that is by the largest aromatic domains for a given particle size and composition. Within the THEMIS framework the largest possible aromatic domain radius (for hydrogen-poor a-C materials) is taken to be half of the particle radius, $a$, that is $a_{\rm R} = a/2$ \citep{2012A&A...542A..98J,2013A&A...558A..62J}.\footnote{Aromatic domains larger than the particle radius lead to a loss of structural coherence because they introduce a weak-bonding, diametric dividing layer.} This requires a corrective term
\begin{equation}
A_{\rm C} = \left( \frac{\pi}{2} \frac{a}{a_{\rm R}} - 1 \right) \frac{e^2}{2a} = \left( \pi - 1 \right) \frac{e^2}{2a},
\end{equation}
in the capacitive energy \citep{LV_2024} 
%to be overcome by the electron 
to IP and EA, 
%that is,  
\[
{\rm IP}(a,Z_{\rm d}) = W_{\rm (s)} + \left( Z_{\rm d} + \frac{1}{2} \right) \frac{\ e^2}{a} + A_{\rm C} 
\]
\begin{equation}
\ \ \ \ \ \ \ \  = W_{\rm (s)} + \frac{\pi}{4} \frac{e^2}{a_{\rm R}} + \frac{Z_{\rm d} \, e^2}{a}
= W_{\rm (s)} + \frac{\pi}{2} \frac{e^2}{a} + \frac{Z_{\rm d} \, e^2}{a},  
\label{eq_IP_charged}
\end{equation}
\[
{\rm EA}(a,Z_{\rm d}) = W_{\rm (s)} - E_{\rm g} +  \left( Z_{\rm d} - \frac{1}{2} \right) \frac{\ e^2}{a} - A_{\rm C}
\]
\begin{equation}
\ \ \ \ \ \ \ \  = W_{\rm (s)} - E_{\rm g} - \frac{\pi}{4} \frac{e^2}{a_{\rm R}} + \frac{Z_{\rm d} \, e^2}{a} 
= W_{\rm (s)} - E_{\rm g} - \frac{\pi}{2} \frac{e^2}{a} + \frac{Z_{\rm d} \, e^2}{a} ,  
\label{eq_EA_charged}
\end{equation}
where $W_{\rm (s)} = 4.6$\,eV is the work function for bulk a-C(:H) materials, $e$ is the electron charge, and $Z_{\rm d}$ is the dust particle charge. In general, in the chemistry and solid-state domains, the electron affinity is defined for a neutral atom, molecule or surface (in our case for a particle with $Z_{\rm d} = 0$). For the neutral state ($Z_{\rm d} = 0$) the above equations for the IP and EA of a seed cluster reduce to    
\begin{equation}
{\rm IP}(a) = W_{\rm (s)} + \frac{\pi}{2} \frac{e^2}{a}, 
\label{eq_IP}
\end{equation}
\begin{equation}
{\rm EA}(a) = W_{\rm (s)} - E_{\rm g} - \frac{\pi}{2} \frac{e^2}{a}. 
\label{eq_EA}  
\end{equation} 
The work function is also size-dependent, $W(a)$, as per the IP(a) trend, and taking the work function for a sphere from \cite{Wong_etal_2003}, as highlighted by \cite{2016MNRAS.459.2751K},  
\begin{equation}
W(a) = W_{\rm (s)} + \frac{1}{4 \pi \varepsilon_0} \frac{3}{8} \frac{e^2}{a} \frac{\epsilon -1}{\epsilon} , 
\label{eq_Wa}
\end{equation}
where $\varepsilon_0$ is the permittivity of free space, and $\epsilon$ is the static dielectric constant \cite[taken to be 7.0 for carbon, following][]{2016MNRAS.459.2751K}. The size-dependent parameters  W, IP, EA, and $E_{\rm g}$ are plotted in Fig.~\ref{fig_params}, along with the IPs for some critical metal atoms (K, Na, Ca, Mg, Fe, Si, and C) and the EA for carbon atoms. This figure illustrates the significant size dependences of $E_{\rm g}$, W, IP, and EA for a-C(:H) nanoparticles with radii $\lesssim 3$\,nm. Notice that for the smallest clusters, and as must be the case as the radius decreases and the number of carbon atoms tends to unity, the cluster IP and the EA approach that for a single carbon atom (see Fig. \ref{fig_params}). 
%This is shown by the red (cluster IP) and blue (cluster EA) lines in (see Fig. \ref{fig_params}), which approach that for atomic carbon at $a \sim 0.5$nm. 
Note that the carbon-carbon bond length is 0.133nm and 0.146nm for C$=$C and C$-$C bonds, respectively, and that a 0.2nm radius cluster therefore contains $\approx (0.2/0.14)^3 \simeq 3$ carbon atoms and a $\sim 0.4$nm radius cluster $\approx [0.4/0.14]^3 = 20-30$ carbon atoms.

%-------------------------------------------------------------------------------------------------------------------------------- 
\subsection{Seed cluster and nanoparticle charging rates}
\label{sect_charging_rates}
%--------------------------------------------------------------------------------------------------------------------------------   
The size-dependence of the adopted \cite {2012A&A...542A..98J}optical properties turns out to have some interesting consequences for the seed cluster an nanoparticle charge distribution. 

A calculation of the grain charge requires the thermal speed of a given species $j$ (an electron $e$, an ion $i$ or a dust particle $d$) of mass $m_j$ and is taken to be the average speed
\begin{equation}
V_j = \left( \frac{8 \, k_{\rm B} \, T_{\rm gas}}{\pi \, m_j} \right)^\frac{1}{2}
\label{eq_vel}
\end{equation}
where $k_{\rm B}$ is the Boltzmann constant and $T_{\rm gas}$ is the gas kinetic temperature. The thermal speed then enters into the cross-sections for the interaction of ions and electrons with charged grains through the Coulomb collision cross-section modifying parameter, $F_{\rm C,j}$. For the collision of a species $j$ with a dust particle $d$ of radius $a_{\rm d}$, the Coulomb factor is 
\begin{equation}
F_{\rm C,j} = 1 - \left( \frac{ 2 \, Z_j \, Z_{\rm d} \, e^2}{a_{\rm d} \, m_j \, V_j^2 } \right) 
= 1 - \left( \frac{\pi }{ 4 } \frac{ Z_j \, Z_{\rm d} \, e^2}{a_{\rm d} \, k_{\rm B} \, T_{\rm gas} } \right),  
\label{eq_FCcoll}
\end{equation}
where 
$Z_j$ and $Z_{\rm d}$ are the charge states ($|Z| \geqslant 0$) of the species $j$ and the seed cluster of nanoparticle $d$. For electrons and ions the relevant collision radius is $a_d$ and where the projectile $j$ is another particle it is $(a_j+a_{\rm d})$. In the following we  consider the principal grain charging processes due to electron and ion collision and sticking, and photoelectron emission. 

The collision rate, $R_{{\rm coll},j}$, between species $j$ (an ion, $i$, or an electron, $e$) and a nanoparticle $d$ is 
\begin{equation}
R_{{\rm coll},j}(a_{\rm d},Z_{\rm d}) = \left( \sigma_{\rm d} \, F_{\rm C,j} \right) \, n_{\rm H} \, X_j \, V_j \, S_j 
\label{eq_coll}
\end{equation}
where $X_j$ is the relative abundance of species $j$, $\sigma_{\rm d} = \pi \, a_{\rm d}^2$ is the particle geometrical collision cross-section,  and $S_j$ is the sticking coefficient. The bracketed term on the right hand side of the above equation is the Coulomb-modified collision cross section. 

The photoelectric emission rate for absorbed photons of sufficient energy to remove an electron from the grain, $R_{\rm pe}$, is 
\begin{equation}
R_{\rm pe}(a_{\rm d},Z_{\rm d},E_{\rm h \nu}) = \sigma_{\rm d} \, Q_{\rm abs}(a_{\rm d},E_{\rm h \nu}) \, F(E_{\rm h \nu}) \, Y_{\rm pe}(a_{\rm d},Z_{\rm d},E_{\rm h \nu}) 
\end{equation}
where $Q_{\rm abs}(a,E_{\rm h \nu})$ is the dust absorption efficiency factor \citep[optical properties taken from][]{2012A&A...540A...1J,2012A&A...540A...2J,2012A&A...542A..98J,2013A&A...558A..62J}, $F(E_{\rm h \nu})$ is the photon flux, and $Y_{\rm pe}(a_{\rm d},Z_{\rm d},E_{\rm h \nu})$ is the dust photoelectron yield for photons of energy $E = {\rm h \nu}$. The carbon dust photoelectron yields (see Appendix A) are calculated using a slightly modified methodology from \cite{2016MNRAS.459.2751K}. 
%The dependency of $R_{\rm pe}$ on the dust charge state is rather more indirect because it enters into the yield $Y_{\rm pe}$ in that a higher positive charge state reduces the photoelectron yield. 

% FIGURE *********************************************************
\begin{figure} 
\centering
 \includegraphics[width=9.5cm]{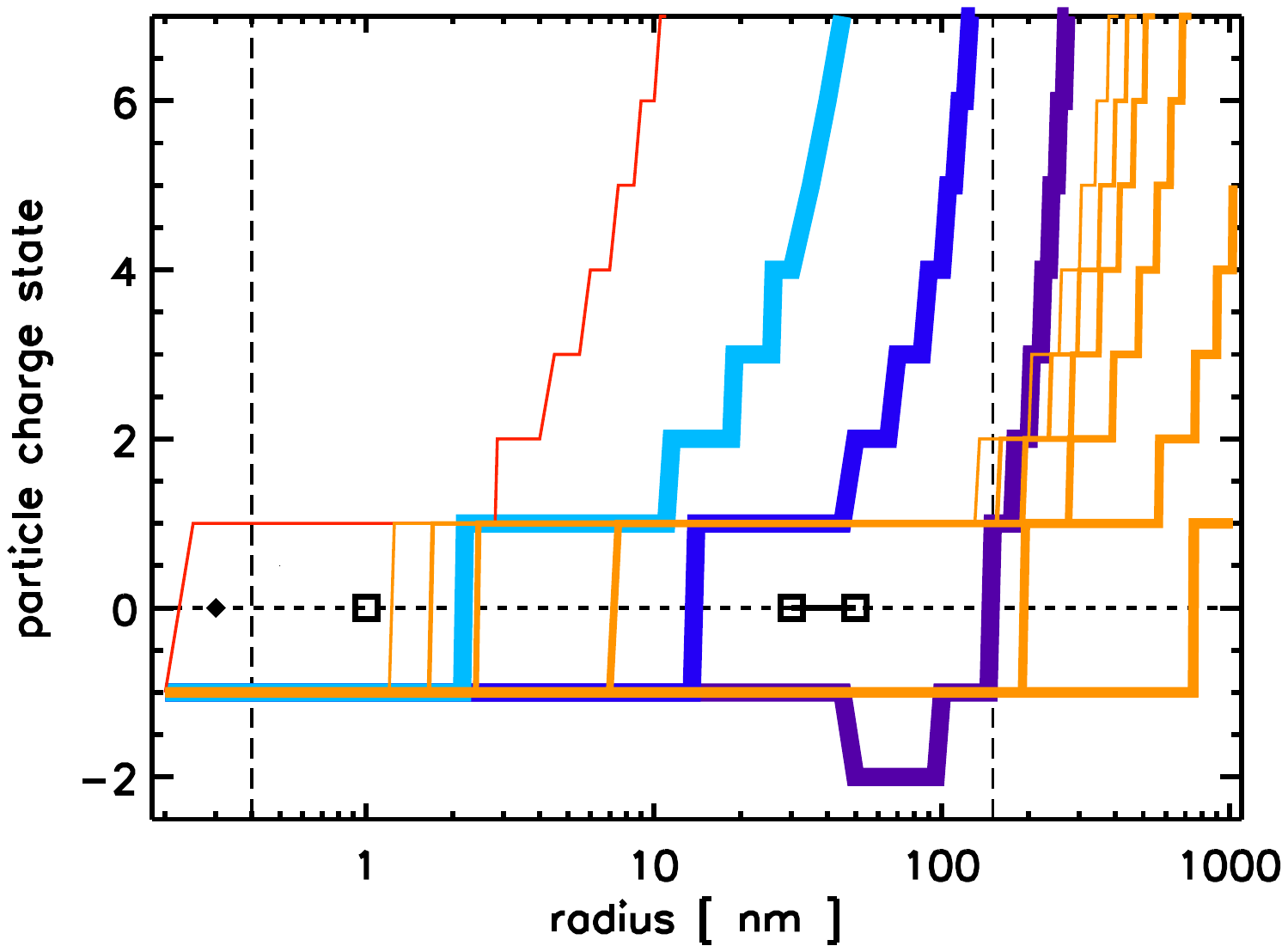}
 \caption{Equilibrium charge states for particles as a function of radius, for the parameter sets indicated in Table \ref{table_conds}
 %, for increasing photon energy and photon flux  but decreasing gas density 
 (purple to blue to cobalt).
 %, for a gas temperature of 1,000K. 
 The squares indicate the derived carbon dust radii, $\sim 1$ and $30-50$\,nm in WR\,140 \citep{2023ApJ...951...89L} and the small diamond marks a seed cluster of radius = 0.3nm. The vertical dashed lines indicate the minimum grain radius and the approximate peak of the large grain size distribution in THEMIS. The red line shows the CNM charge states and the orange lines those for dense cloud edges with $A_{\rm V}$ of  0, 0.25, 0.5, 1, 2, and 3 (thin to thick orange lines, respectively).} 
\label{fig_charge}
\end{figure}
%  *********************************************************

The equilibrium seed cluster and nanpoparticle charge states were determined by solution of the following equation for $Z_{\rm d}$ 
\begin{equation}
0 = - R_{{\rm coll},e}(a_{\rm d},Z_{\rm d}) + R_{{\rm coll},i}(a_{\rm d},Z_{\rm d}) + R_{\rm pe}(a_{\rm d},Z_{\rm d},E_{\rm h \nu}). 
\end{equation}
It is usual to solve this kind of equation for the grain surface potential, that is $\phi = Q/a_{\rm d} = ( Z_{\rm d} e )/a_{\rm d}$. However, given that this is directly proportional to the number of charges on the particle, $Z_{\rm d}$, we prefer to directly solve this equation for the nearest integer values of the grain charge, that is for the quantised charge states of the particles. While this approach does not resolve the charge distribution for a given size grain it does indicate the most probable charge state for a given size of grain. The equilibrium charge states are show in Fig. \ref{fig_charge} for dust as a function of radius for the photon energies and fluxes, gas densities and temperatures shown in Table \ref{table_conds}, for incident Ca$^+$ ions with an assumed relative abundance $X$(Ca$^+$) $= 10^{-5}$  
%, C, O, and electrons.  
%and 0.2nm seed clusters interacting with nanoparticles of radii from 0.2 to 3nm at a gas temperature of 1000K and density of $2 \times 10^7$cm$^{-3}$. 
%The assumed abundances, relative to hydrogen, $X$(Ca$^+$), $X$(C), $X$(O), 
%and for the 0.2nm seed clusters, 
%are $10^{-5}$, $2 \times 10^{-3}$, $4 \times 10^{-4}$, 
%$0.1X$(C)$/N_{\rm C}(0.2{\rm nm})$, 
%respectively,\footnote{The elemental abundances for C and O were guided by the CO nova outburst modelling of \cite{2016A&A...593A..54J}.}
% and it was assumed that 10\% of the carbon atoms are in the form of 0.2nm carbon clusters.} 
%where $N_{\rm C}(0.2{\rm nm}) \simeq 2-3$ is the number of C atoms in the cluster, 
and $X({{\rm e}^-})$ taken to be equal to $X$(Ca$^+$). Some key values of the Coulomb factors for each of the considered sets of physical conditions are listed in Table \ref{table_F_Coulomb}. One thing of particular note that can be gleaned from Fig. \ref{fig_charge} and Table \ref{table_F_Coulomb} is that the critical radius for  disproportionation, where the particle charge switches sign, increases with increasing gas density and extinction.

The reason for the anionic charge states of the seed clusters ($a \simeq 0.2 - 0.4$nm) and nanoparticles is because their photoelectron yields are essentially zero for photon energies $< 6$\,eV. For grain radii of 1, 10 and 100nm the yields are $\lesssim 0.2$, $\lesssim 0.05$ and $\lesssim 0.009$, respectively \citep[e.g.,][see Fig. \ref{fig_Ype}]{2016MNRAS.459.2751K}.\footnote{Clusters of radii $02-0.4$nm contain a few up to $\simeq 30$ carbon atoms.} 
Further, the photon absorption efficiencies, $Q_{\rm abs}(a,E_{\rm h \nu})$, of grains smaller than 30nm for photon energies less than 6eV are low \citep[$\simeq 10^{-2}$,][]{2012A&A...540A...1J,2012A&A...540A...2J,2012A&A...542A..98J,2013A&A...558A..62J}. Thus, small C$_n$ clusters and nanograins charge negatively because they have high electron affinities, are inefficient photon absorbers, and do not emit photoelectrons; aided by the frequent electron collisions. For larger particles ($a \gtrsim 100$\,nm) these conditions no longer hold and they tend to charge positively through the effects of photoelectric emission. 
%In the following we explore the consequences of these results. 

% FIGURE *********************************************************
\begin{figure*} 
\centering
 \includegraphics[width=18.5cm]{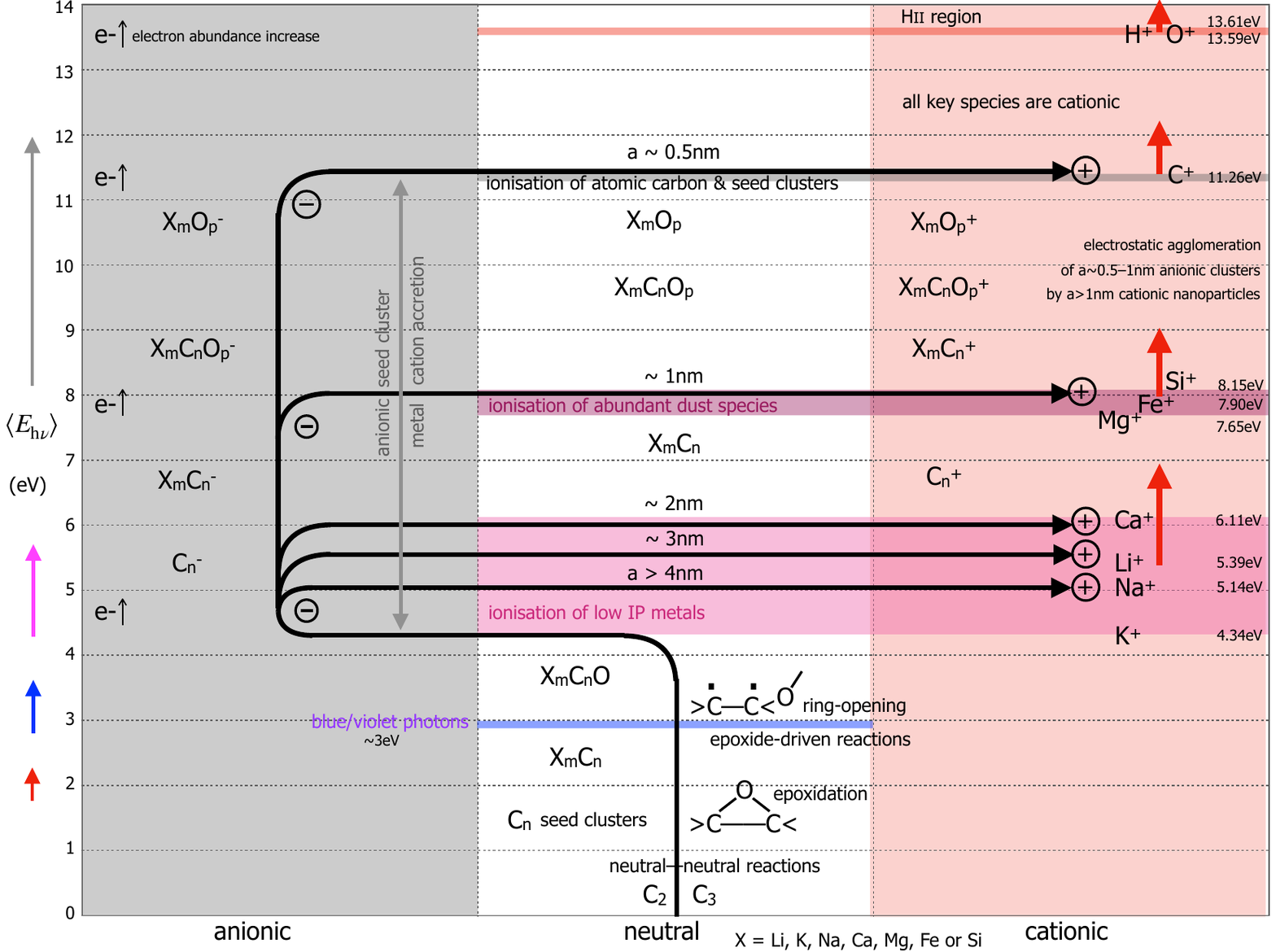}
 \caption{Schematic view of seed cluster formation, charge states and evolution (black lines, see Fig. \ref{fig_params} for seed cluster IPs) during dust nucleation in a hardening radiation field (bottom to top). The ionisation stages (K|Na|Li|Ca, Mg|Fe|Si, C, and H|O), neutral-neutral reactions, epoxidation, epoxide ring opening-driven chemistry, cation accretion and electrostatic agglomeration are indicated. For simplicity hydrogen atoms are not indicated.} 
\label{fig_key}
\end{figure*}
%  *********************************************************

%-------------------------------------------------------------------------------------------------------------------------------- 
\section{A scenario for seeding dust nucleation}
\label{sect_seeding}
%-------------------------------------------------------------------------------------------------------------------------------- 

This section is an attempt to weave together the ideas and processes presented in the previous two sections into a coherent 
% picture that encompasses a possible 
scenario for seeding dust nucleation in dust-forming regions, with the particular emphasis on 
%aim of elucidating these processes as they might occur in 
the rapid dust-forming environments of nov\ae, WR stars, and LBV systems. 

To this aim the reactions described in Sections \ref{sect_nucleation_a} and \ref{sect_nucleation_b} (Eqns. \ref{eq_scheme_1a} to \ref{eq_scheme_2b}) were redeveloped into Eqn. (\ref{eq_sequence}), where for simplicity the bonds to H atoms have been ignored. 
\[
[1]\ {\rm C}_{n} + \ {\rm C} \hspace{1.3cm} \rightarrow \ {\rm C}_{n+1} \hspace{1.9cm} {\rm neutral \ chemistry,}
\]
\[
[2]\ X_{m}{\rm C}_{n} + \ {\rm O} \ \ \ \ \ \ \ \ \ \, \rightarrow \ X_mC_{n-2}({\rm C}- \hspace{-0.3cm}^{\rm O}\,{\rm C}) \hspace{0.45cm} {\rm epoxidation,}
\]
\[
[3]\ X_mC_{n-2}({\rm C}-\hspace{-0.3cm}^{\rm O}\,{\rm C}) \ \ \, \rightarrow \hspace{-0.4cm}^{{\rm h}\nu} \ \ \ X_{m-1}{\rm C}_{n} \ + \ X{\rm O} \ \ \ \ \, {\rm epoxide \ reaction,}
\] 
\[
[4]\ X \ + \ {\rm h}\nu \ \ \ \ \ \ \ \ \ \ \ \ \ \, \rightarrow \ X^+ + \ e^-  \\\\\ \ \, {\rm metal \ ionisation,}
\]
\[
[5]\ X_m{\rm C}_n \ + \ e^- \ \ \ \ \ \ \ \rightarrow \ X_m{\rm C}_n^- \hspace{1.8cm} {\rm seed \ anionisation,}
%m \geqslant 0, \ n \geqslant 2,
\]
\[
[6]\ X_m{\rm C}_n^- \ + \ {\rm C} \ \ \ \ \ \ \ \, \rightarrow \ X_m{\rm C}_{n+1}^- \hspace{1.5cm} {\rm C \ dust \ growth,}
\]
\[
[7]\ X_m{\rm C}_{n}^- \ + \ X^+ \ \ \ \ \ \, \rightarrow \ X_{m+1}{\rm C}_{n} \\\\\ {\rm metal \ accretion,}
\]
%\[
%[6]\ X_mC_{n-2}({\rm C}-\hspace{-0.3cm}^{\rm O}\,{\rm C})^{-} \rightarrow\hspace{-0.4cm}^{{\rm h}\nu} \ \ \ X_{m-1}{\rm C}_{n} \ + \ X{\rm O}^{-} \ \ \ \, {\rm reduction, \ and}
%\]
\[
[8]\ X_m{\rm C}_{n}{\rm O}^{-} \hspace{1.1cm} \rightarrow \hspace{-0.4cm}^{{\rm h}\nu} \ \ \ X_{m-1}{\rm C}_{n} \ + \ X{\rm O}^{-} \ \ \ \, {\rm reduction, \ and}
\]
\begin{equation}
[9]\ X_{m+1}{\rm C}_{n}^{-} + \ {\rm O} \ \ \ \ \, \rightarrow \ X_m{\rm C}_{n} + \ X{\rm O}^- \hspace{0.75cm} {\rm O \ dust \ growth}, 
\label{eq_sequence}
\end{equation}
where $n \geqslant 2$ and $m \geqslant 0$. The processes are shown in the approximate order of their sequential occurrence and  
 are referred to by a prefixed equation sub-number, [N], in the following discussion. Note that metal accretion [7] is effectively driven by charge disproportionation, arising from the interaction of reduced ($X_m{\rm C}_{n}^-$) and oxidised ($X^+$) species, and is  Coulomb-enhanced. Apart from the metal ionisation [4] and seed cluster reduction [8], all of the other processes are charge mediated or induced-dipole interactions \citep[e.g.,][]{Shore:2004hj}. 

Before metal ionisation furnishes free electrons to the gas only neutral-neutral chemical reactions are possible [1], 
which will be accompanied by the ready epoxidation of C$=$C bonds in carbonaceous clusters by incident O atoms [2]. 
The formed epoxide functional groups provide a latent catalytic role by stocking bond-forming potential that will be released by photon irradiation and/or by atomic interactions. Thus, as the radiation field hardens blue-violet photons will initiate epoxide ring-opening and lead to the formation new bonds with incident atoms [3]. 
As K, Na, Li and/or Ca ionisation is initiated by $4.34-6.11$eV UV photons [4]  the growing X$_m$C$_n$ seed clusters will sweep up the furnished electrons rendering them anionic [5], 
X$_m$C$_n^-$, and they can grow by accreting C and O atoms [6,9] and cations [7]. 
The latter metal cation accretion reaction [7] will be enhanced by a Coulomb factor of $\sim 170$ for Ca$^+$ ions incident on anionic nucleating clusters of radius 0.2nm at a gas temperature of 1000K (see Appendix \ref{sect_novae}). 
As the radiation field hardens to $\langle E_{\rm h \nu} \rangle = 7.65-8.15$eV the accretion of K, Na, Li, and Ca cations will be overshadowed by the more abundant Mg, Fe, and Si cations. 
Incident electronegative O atoms will at some stage reduce cluster anions by relieving them of their electrons and may in the process combine with metal atoms (Mg, Fe, and Si) in the seed cluster to form anionic metal oxide, $X$O$^-$, seed clusters [8]. 
%, paving the way for the co-spatial and co-temporal formation of distinct carbonaceous and silicate dusts. 
The final growth reaction in the above sequence [9] re-generates neutral clusters [2] and highlights the cyclic nature of the metal-carbon cluster catalytic nucleation process. It also underlines a likely chemical kinetic scenario for the co-spatial and co-temporal formation of carbon-rich and oxygen-rich dust species under dust nucleating conditions [6,7,9]. 

The above reaction scenario is shown schematically in Fig. \ref{fig_key}, which perhaps better illustrates the sequence of the processes as the stellar radiation field hardens (from bottom to top). This figure emphasises that, prior to the ionisation of low IP metals, the chemistry is driven by neutral-neutral reactions and the likely stocking of latent bond formation energy in epoxide rings. Under blue-violet photon irradiation the epoxide rings will open up and drive the chemistry via new bond formation with incident species. 
After low IP metal ionisation ($\sim 4-6$eV)) and before C atom ionisation (11.26eV) anionic seed clusters form and can react with metal cations (K$^+$, Na$^+$, Li$^+$, Ca$^+$, Mg$^+$, Fe$^+$, and Si$^+$) and also with abundant atoms (e.g. C, O, and N). Note that the upper size limit of the anionic seed clusters decreases as the radiation field hardens because increasingly larger nanoparticles are progressively ionised. The minimum possible radius for anion seed clusters will always be $\approx 0.5$nm but as the mean photon energy, $\langle E_{\rm h \nu} \rangle$, of the radiation field increases their upper size limit approaches this minimum until all seed clusters and dust nanoparticles are positively charged (see Fig. \ref{fig_params} for size-dependence of the seed cluster IPs). 

Circumstantial evidence for the types products proposed in the above dust seeding scenario might perhaps be found in the observation of carbon-bearing anionic \cite[e.g. C$_n^-$ and C$_n$H$^-$, $n=7-23$,][]{2000MNRAS.316..195M} and Mg-bearing organometallic \cite[e.g. MgC$_n$H, $n=2-4$,][]{2019A&A...630L...2C} chains in the circumstellar shell of the carbon star IRC+10216. These linear species indicate a viable route to the formation of similar chemical species in circumstellar regions.  

In nov\ae\ the rapid growth of dust is associated with the onset of UV transparency, in the declining optical light curve, and is clearly triggered immediately before the optical decline towards the minimum due to the opacity of newly formed dust. This is the so-called dust dip which correlates with the excess in the infra-red emission arising from that same dust. After this epoch of dust formation the opacity reaches a maximum before decreasing due to dilution in the expanding outflow  \citep{Shore:2004hj,Shore:2018bi}. 

It therefore appears to be an inescapable conclusion that photo-initiated processes must play a pivotal role in dust formation, given that nova dust formation occurs in association with the onset of UV transparency \citep[e.g.][]{Shore:2018bi}. Principal among these processes is the ionisation of low IP metals (e.g. K, Na, Li, and Ca) as the radiation field hardens [4]. These ions and the liberated electrons can then trigger a series of reactions (anionic cluster formation [5] and metal accretion [7]) that result in the formation of seed clusters that initiate dust nucleation and its subsequent growth through ion and atom accretion [6,7,9].

For the nov\ae\ and WR system environmental conditions proposed in Table \ref{table_conds} by far the most frequent collision partners with the seed clusters ($0.2-0.3$nm, with $\simeq 3-10$ carbon atoms) are electrons [5] (see Fig. \ref{fig_coll}). Carbon atoms are the most frequent atomic collision partners [1,6], closely followed by Ca$^+$ ions [7] aided by the large Coulomb factor ($\gtrsim 100$) which counteracts the low Ca atom abundance ($10^{-5}$). Oxygen atom collisions [2,9] ($X$(O) $= 8 \times 10^{-4}$) with seed clusters  are almost as frequent as Ca$^+$ collisions [7]. The carbonaceous dust-nucleating seeds interact with a Ca$^+$ ions and C and O atoms [1,2,6,7,9] over timescales of the order of days to a few weeks at the onset of ionisation. 

From Fig. \ref{fig_charge} and Table \ref{table_conds} it can be seen that the disproportionation radius decreases with decreasing gas density in the nov\ae\ and WR cases. In these environments the metals K, Na, Li, and Ca will ionise [4] as the gas becomes transparent in the UV (i.e. $\langle E_{ \rm h \nu} \rangle \sim 6$eV) and this will in turn drive the appearance of anionic seed clusters [5] and lead to the appearance of disproportionation, triggering dust nucleation via anionic seed cluster accretion onto larger positively charged particles. However, as the density in the outflowing gas decreases the disproportionation radius will shift to smaller radii until a point where it can no longer be sustained and all grains will be positively charged. As Fig. \ref{fig_key} indicates this seemingly occurs once the mean photon energy of the radiation field is sufficient to ionise C atoms (i.e. $\langle E_{ \rm h \nu} \rangle \geqslant 11.26$eV). Within this scenario there would therefore seem to be a somewhat restricted window, a sweet spot, for rapid dust nucleation and growth, with timescales seemingly of the order of days to weeks in dense circumstellar environments ($n_{\rm H} \simeq 10^7 - 10^8$cm$^{-3}$). 

Given the uncertain details of the reaction pathways and their associated rates it will not be easy to implement the above scenario into a meaningful dust nucleation model beyond the initial atomic, ionic and electron interactions. However, this will be possible once the reactions of mixed carbon, metal and oxygen polyatomic neutral and charged cluster species containing up to a hundred atoms have been measured in the laboratory or been calculated quantum mechanically.

%\begin{color}{blue} 
%\noindent 1) Expand the above to give a more complete vision of the dust seeding scheme and processes proposed here. \\ 
%2) Shorten charge section results application and/or move stuff to an appendix. \\
%3) Add new section after that combines ans summarises the dust seeding processes of low IP metal ionisation, ion-cluster reactions, de-epoxidation leading to C-rich and O-rich dusts. \\ 
%\end{color}

Laboratory evidence has uncovered a nova grain containing silicate and oxide nanocrystalline inclusions that formed before the host graphite and in the same nova ejecta \citep{2019NatAs...3..626H}. This observation is inconsistent with thermodynamic equilibrium dust condensation in nov\ae\ and therefore points towards kinetic nucleation and condensation processes in dust formation, much as those proposed and studied here. This important laboratory result suggests that the C/O ratio may therefore be of rather limited value in determining the dust composition during nucleation and condensation \citep{2019NatAs...3..626H}. 

%In considering the gas-dust interaction in astrophysical media it is useful to consider an inverse accretion rate scaling parameter proportional to $( n_{\rm H} \surd T_k$) as proposed by \cite{2018arXiv180410628J}. For example, the high density phase of the diffuse neutral ISM contains a cold neutral medium (CNM) with kinetic temperature $T \sim 3-200$K and  density $n_{\rm H} \sim 5-120$cm$^{-3}$ \cite[e.g.][]{1993AJ....106.2349L,2003ApJ...587..278W} and taking $n_{\rm H} \simeq 10^2$cm$^{-3}$ and $T_k \simeq 10^2$K this gives $n_{\rm H} \surd T_k  = 10^3$. In the denser media of nov\ae\ and WR circumstellar environments, with $n_{\rm H} \simeq 10^8$cm$^{-3}$ and $T_k \simeq 10^3$K, $n_{\rm H} \surd T_k  = 3 \times 10^9$ and the gas-dust interactions will be more than million times faster.  Note that no Coulomb effects have been taken into account here. Thus, what would occur over time scales of $\sim 10^7$\,yr in the CNM  may take only a matter of days in the dense circumstellar media of nov\ae\ and WR systems.  

%This section has set the scene for what follows, in the sense that 
%The formation of critical clusters, stable entities that provide the seeds for dust nucleation, is at the heart of the dust formation and growth processes, of which the size-dependent particle charge distribution is a key and pivotal element. 

%\begin{color}{blue}
%Bring all the above together into a coherent picture for the initial (seeding) phases leading to the formation of critical clusters and hence to dust nucleation. 
%\end{color} 

Appendix \ref{sect_novae} presents a brief discussion of how dust growth might proceed in nov\ae\, and in WR and LBV systems.

%-------------------------------------------------------------------------------------------------------------------------------- 
\subsection{Nanoparticle accretion in interstellar media}
\label{sect_ISM_growth}
%-------------------------------------------------------------------------------------------------------------------------------- 

%\begin{color}{blue}
This section briefly considers how the above discussion may apply to ISM cloud studies. 
The dust charge calculations for the CNM and cloud cases (see Fig. \ref{fig_charge} and Table \ref{table_conds}) assume a mean photon energy of 7eV, a photon flux of $3 \times 10^7$photons cm$^{-2}$ s$^{-1}$ \citep[e.g.][]{2002ApJ...570..697H}, and typical gas densities (100cm$^{-3}$) and temperatures (100K). The cloud dust charge calculations assume the same photon energy, a slightly reduced photon flux of $10^7$photons cm$^{-2}$ s$^{-1}$, a cloud density of 2,000cm$^{-3}$, and a gas temperature of 20K. For the cloud case the photon flux is attenuated by a factor $e^{-A_{\rm V}}$ for $A_{\rm V} =0$, 0.25, 0.5, 1, 2, and 3. 

As indicated in Fig. \ref{fig_charge}, for $A_{\rm V} < 1$ only sub-7nm radius grains are negative, and for  $A_{\rm V} \geqslant 2$ all grains smaller than 200nm are negatively charged. Note that in the cloud case the charge disproportionation radius increases with increasing extinction. For the particular case of $A_{\rm V} = 1$ all grains with radii less than 7nm, essentially all of the carbonaceous a-C nanoparticles in the THEMIS model \citep{2013A&A...558A..62J,2017A&A...602A..46J,2014A&A...565L...9K,2024A&A...684A..34Y}, are negatively charged and all larger grains are positively charged. As a result of charge this disproportionation where  
%$\delta_\pm |Z_{\rm d}| = 2$ and 
the sign of the grain charge flips at $a = 7$nm, the collision rates between smaller and larger grains may be enhanced by a factor of up to 25 for collisions between nanoparticles ($a = 0.4-1$nm) and 5nm radius grains and a factor of 2 for their collision with 150nm radius grains. %For $A_{\rm V} = 0$ the charge disproportionation for $a = 0.4$ to 1nm ($Z = -1$) in collision with $a = 150$nm grains ($Z = +2$) leads to a Coulomb enhancement factor of $\sim 3$. 
%\end{color} 

Appendix \ref{sect_ISM} adds a little to this discussion of enhanced nanoparticle accretion in the CNM of the ISM.

%-------------------------------------------------------------------------------------------------------------------------------- 
\section{Discussion and conclusions}
\label{sect_conclusions}
%-------------------------------------------------------------------------------------------------------------------------------- 

This work reconsidered the fundamental processes of dust nucleation in terms of the likely chemical nature and interactions of carbonaceous seed clusters, containing perhaps only tens of atoms, and their interactions with ions, atoms and other nanoparticles. In particular the critical process of O atom reactions with C$=$C bonds to form a three atom C$_2$O epoxide rings (epoxidation) and the processes that drive the formation of negatively charged seed clusters and positively charged larger nanoparticles (disproportionation) were investigated. 

The dust seeding scenario presented here proposes previously unexplored mechanisms whereby both carbon-rich and oxygen-rich dust species may form contemporaneously and co-spatially. 

Under the specific conditions of dust formation in nov\ae, WR, and LBV systems, where a rapidly changing  radiation field is associated with dust formation, it is possible that seed cluster epoxidation and dust charge disproportionation could play critical roles. These newly considered processes could drive an  increase in the dust growth rates through photon-initiated epoxide ring-opening reactions leading to new bond formation with incident atoms and the Coulomb enhancement of collisions between negatively-charged seed clusters 
%and nanoparticles (dust nuclei) 
and the positively-charged larger particles. These  processes are particularly important for carbonaceous grain nucleation and growth but could also drive O-rich dust formation through incident O atom reactions. Given that carbon and silicate nanoparticles have similar physical properties, such as IP, EA and UV photon absorption efficiencies
\cite[e.g.][]{2006ApJ...645.1188W,2016MNRAS.459.2751K}, it is likely that these same effects will also be important for the nucleation and growth of oxygen-rich dust species. 

Under weakly ionising conditions ($E_{\rm h \nu} \lesssim 7$\,eV) it is possible that dust nucleation could be driven by low ionisation potential metals such as K, Na, Li, and Ca (all with IP $\lesssim 6$\,eV).  
Once low IP alkali metal and alkaline earth metals such as K, Na, Li, and Ca are ionised their collision with anionic seed clusters is Coulomb-enhanced. For example, at typical nova gas temperatures and densities, the collision of Ca$^+$ ions with seed clusters may be increased by orders of magnitude, thus counteracting the low abundance of these low IP metals. 

Although the charge disproportionation particle collision enhancement factors of $\sim 3-5$ in dust nucleating regions may be relatively modest, compared to more than an order of magnitude in the outer regions of molecular clouds ($n_{\rm H} \sim$ few $\times 10^3$), principally due to the order of magnitude differences in their typical gas temperatures, the considerably higher density in the former regions ($n_{\rm H} \sim 10^7 - 10^8$) is a significant contributing factor to rapid dust formation. 

It appears that grain growth via the accretion of nano-particles onto larger grains in the outer regions of interstellar clouds should be enhanced by charge disproportionation effects, and possibly by larger factors than in dust nucleating regions because of the lower gas phase temperatures. This may help to explain the rapid disappearance of (sub)nanometre dust species via accretion onto micro-sized grains in the transition from the low density CNM to dense molecular clouds and circumstellar discs. 

The enhanced dust growth effect explored here clearly needs to be investigated through more detailed and targeted studies of dust formation under the particular conditions presented by the unusual nova, WR and LBV dust-forming systems and dust-growth regions in interstellar clouds. This preliminary study appears to offer some help in explaining the fast dust nucleation and growth observed in nov\ae, WR and LBV systems, and also the enhanced nanoparticle accretion in the outer regions of dense interstellar clouds.

%\begin{acknowledgements}
%The author wishes to thank Laurent Verstraete and Nathalie Ysard for their critical input into earlier incarnations of this study, and the referee for a thorough and detailed review. 
%\end{acknowledgements}

\bibliographystyle{aa} 
\bibliography{Ant_bibliography.bib} 

\begin{appendix}

%-------------------------------------------------------------------------------------------------------------------------------- 
\section{Photoelectric emission}
\label{sect_PE_params}
%-------------------------------------------------------------------------------------------------------------------------------- 

\begin{figure}[h] 
\centering
 \includegraphics[width=9.5cm]{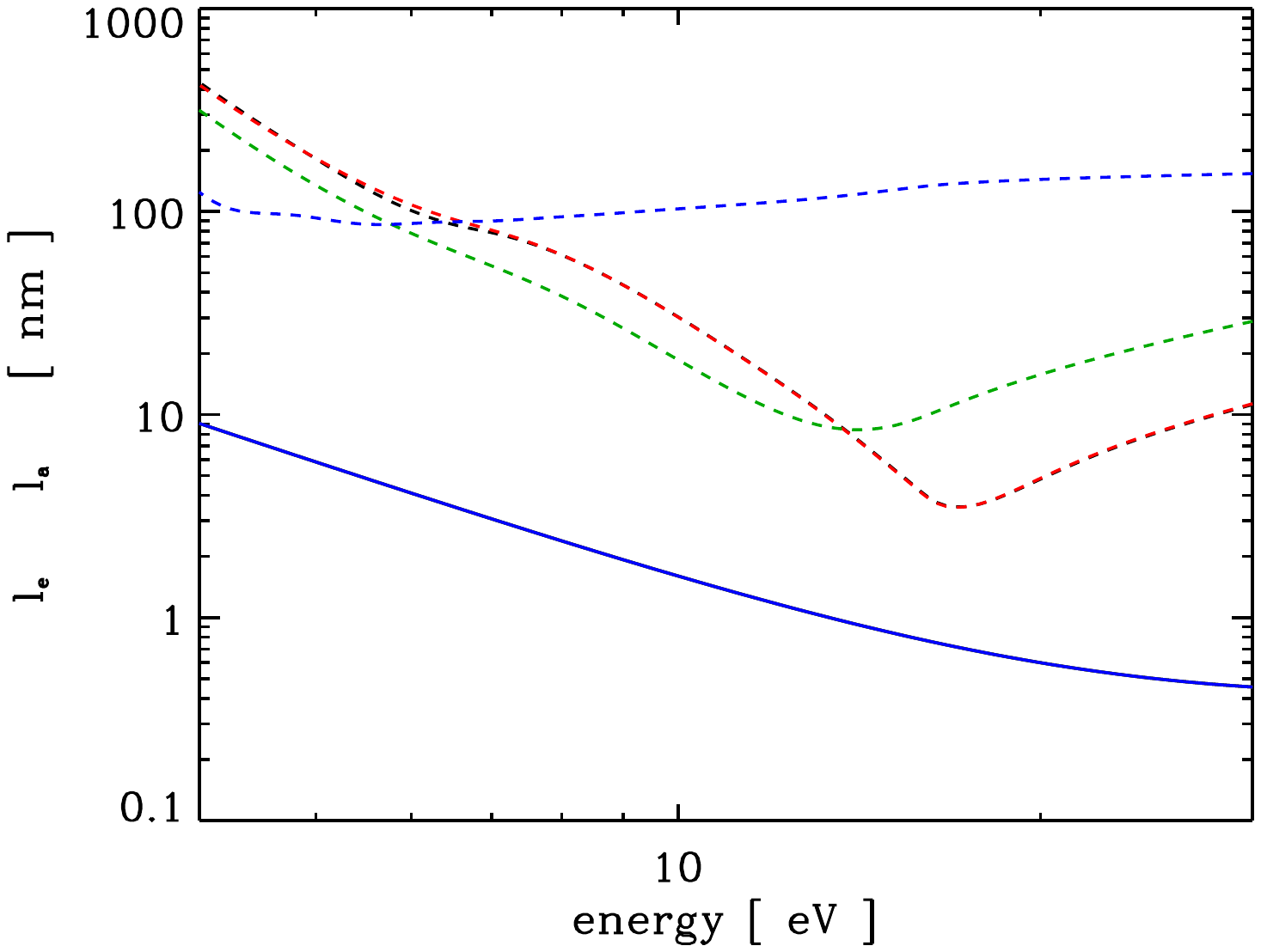}
 \caption{The photon attenuation lengths (dashed lines) for spherical a-C(:H) grains of radii 0.37 (black), 1 (red), 10 (green), and 100\,nm (blue). The solid blue line shows the electron mean free path length as a function of energy, which is the same for all grain sizes.}
\label{fig_lale}
\end{figure}

\begin{figure}[h] 
\centering
 \includegraphics[width=9.5cm]{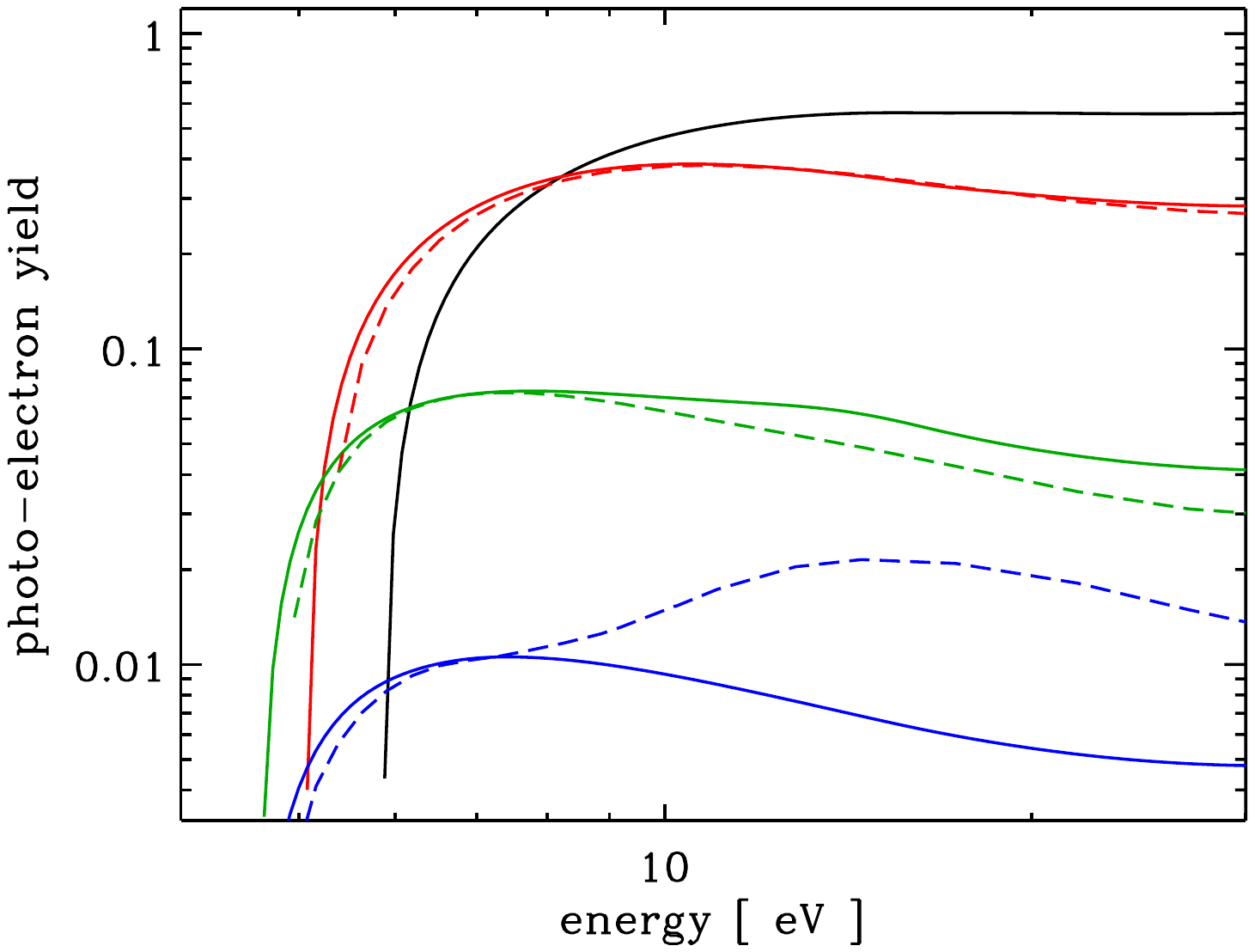}
 \caption{The photoelectric yields (solid lines) for spherical a-C(:H) grains of radii 0.37 (black), 1 (red), 10 (green), and 100\,nm (blue).  The dashed lines show the photoelectric yields tabulated in \citep{2016MNRAS.459.2751K} for 1, 10, and 100\,nm carbon grains.}
\label{fig_Ype}
\end{figure}

We followed the approach developed by \cite{2016MNRAS.459.2751K} but have instead adopted $W_\infty = 4.6$\,eV and define the photon attenuation length, $l_a$, as 
\begin{equation}
l_a = \frac{4 a}{3 Q_{\rm abs}}.
\end{equation}
We otherwise use Eq. (A12), and all the other related equations and parameters, from \cite{2016MNRAS.459.2751K} to derive the size-dependent photoelectric yields. In Fig. \ref{fig_lale} we show our derived photon attenuation and electron mean free path lengths and in Fig. \ref{fig_Ype} our photoelectric yields compared to those tabulated by \cite{2016MNRAS.459.2751K} based on the adopted methodology.

% FIGURE *********************************************************
%  *********************************************************

% FIGURE *********************************************************
%  *********************************************************

%-------------------------------------------------------------------------------------------------------------------------------- 
\section{Dust growth}
\label{sect_growth}
%-------------------------------------------------------------------------------------------------------------------------------- 

Here we consider the collision rate between two particles with a view to exploring grain growth in dust-forming regions and in the outskirts of interstellar clouds in the CNM.

Depending on the environment the interacting particles may be dust nucleating clusters, small or large grains. The collision rate between a particle of radius $a_1$ and charge $Z_1$ with another of radius $a_2$ and charge $Z_2$ is given by 
\begin{equation}
R_{{\rm coll},dd}(Z_1,Z_2) = n_{\rm H} \ X_1 \ \pi \, (a_1 + a_2)^2 \ V_1 \ S_1 \ F_{\rm C,dd}(Z_1,Z_2). 
\end{equation}
From  Eq. (\ref{eq_vel}) $V_1$ can be expressed as $0.042 \surd(T_{\rm gas}/n_{\rm C})$km/s, where $n_{\rm C}$ is the number of carbon atoms in the particle. 
For charged grain interactions, the collision cross-section is modified by the Coulomb factor 
\begin{equation}
F_{\rm C,dd}(Z_1,Z_2) = 1 - \left( \frac{ 2 \ Z_1 \, Z_2 \ e^2}{(a_1+a_2) \ m_1 \ V_1^2 } \right).  
\end{equation}
Where the clusters and grains are oppositely charged this will lead to an enhancement of the grain-grain collision rates. A charge disproportionation parameter $\delta_{\pm} |Z_{1,2}|$, defined as the sum of the offsets from zero of the opposite charges on the interacting particles $\delta_{\pm} |Z_{1,2}| = |Z_1|+|Z_2|$, is used in the following: the minimum value for disproportionation is then $\delta_{\pm} |Z_{1,2}| = 2$.  
For particle Brownian motion speeds ($\sim 0.2-1$km s$^{-1}$ for nov\ae\ dust nucleation and $\sim 0.05-0.25$km s$^{-1}$ for CNM cloud conditions for $n_{\rm C} = 30$ to 3, respectively), where $\delta_{\pm} |Z_{1,2}| = 2$, the collision rate between singly charged anionic ($Z = -1$) and cationic ($Z = +1$) particles can be enhanced by factors of $3-5$, for $T_{\rm gas} = 1000$K, and factors of $2-25$, for $T_{\rm gas} = 100$K, as indicated in Table \ref{table_F_Coulomb}. Thus, the effect of disproportionation is to enhance grain-grain collision rates and, if the grains stick together upon collision, will lead to enhanced dust growth rates.  

\begin{figure} 
\centering
 \includegraphics[width=10cm]{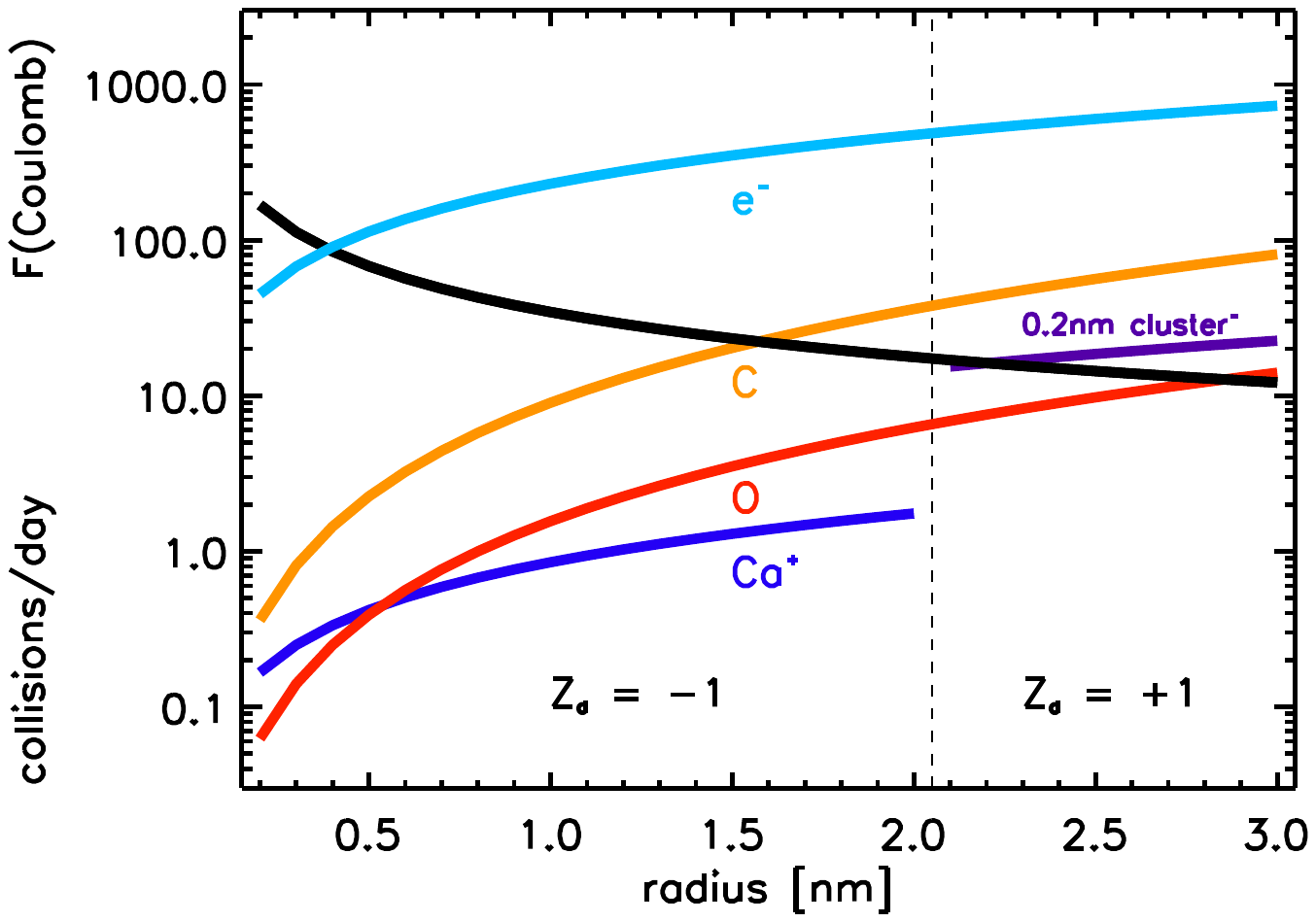}
 \caption{Coulomb factor as a function of radius for collisions between Ca$^+$ ($Z=+1$) and negatively charged grains ($Z=-1$,  black line).  The cobalt line gives the electron collision rate with neutral grains ($Z=0$).  Number of collisions per day of Ca$^+$ (with anionic particles $a < 2$nm, blue), C atoms (orange), and O atoms (red) as a function of the target particle radius. The vertical dashed line indicates a disproportionation radius at 2.05nm. Also shown are the number of collisions per day of 0.2nm clusters for target particle radii $>2$nm (purple).} 
\label{fig_coll}
\end{figure}

Some illustrative Coulomb collision factors and collision timescales (in collisions day$^{-1}$) for dust nucleating regions, derived from Eqns. (\ref{eq_vel}) to (\ref{eq_coll}), are shown in Fig. \ref{fig_coll} 
%% 
%\\ FOLLOWING ALREADY USED ABOVE IN MAIN BODY SO NEED TO CUT HERE. \\
for gas phase species (Ca$^+$, C, O, and electrons) and 0.2nm seed clusters interacting with nanoparticles of radii from 0.2 to 3nm at a gas temperature of 1000K and density of $2 \times 10^7$cm$^{-3}$. The assumed abundances, relative to hydrogen, $X$(Ca$^+$), $X$(O), $X$(C),\footnote{The elemental abundances for C and O were guided by the CO nova outburst modelling of \cite{2016A&A...593A..54J}.} and for the 0.2nm seed clusters, are $10^{-5}$, $2 \times 10^{-3}$, $4 \times 10^{-4}$, $0.1X$(C)$/N_{\rm C}(0.2{\rm nm})$, respectively, where $N_{\rm C}(0.2{\rm nm}) \simeq 2-3$ is the number of C atoms in the cluster, and $X({{\rm e}^-}) = X$(Ca$^+$).
%\footnote{The elemental abundances for C and O were guided by the CO nova outburst modelling of \cite{2016A&A...593A..54J} and it was assumed that 10\% of the carbon atoms are in the form of 0.2nm carbon clusters.} 

% FIGURE *********************************************************
%  *********************************************************

The principle processes driving charge disproportionation around the neutrality condition $\delta_{\pm} |Z_{1,2}| \geqslant 2$ are those that tend to favour the anionic states of clusters and small grains ($n_{\rm C} < 100$) in a given environment (see Fig. \ref{fig_charge}). See the end of Section \ref{sect_nucleation_b} for more details. 
%The critical characteristics are low photoelectron yields and low photon absorption efficiencies but high electron affinities. For example, the cluster  photoelectron yields are essentially zero for photon energies $< 6$\,eV \citep[e.g. see Fig. \ref{fig_Ype},][]{2016MNRAS.459.2751K}, the photon absorption efficiencies, $Q_{\rm abs}(a,E_{\rm h \nu})$, for this same photon energy range are $\simeq 10^{-2}$ \citep{2012A&A...540A...1J,2012A&A...540A...2J,2012A&A...542A..98J,2013A&A...558A..62J}, and the electron affinities are significant ($\simeq 1-2$eV). All of these effects conspire to reinforce the anionic charge states of the smallest clusters ($a \simeq 0.2 - 0.4$nm) and are further aided by the frequent electrons. 

%-------------------------------------------------------------------------------------------------------------------------------- 
\subsection{Dust formation in nov\ae}
\label{sect_novae}
%-------------------------------------------------------------------------------------------------------------------------------- 

Nova-formed grains appear to be large \citep[radii $\geqslant0.1\mu$m, e.g.][]{Shore:2018bi} and to form on timescales of the order of  weeks to months. The dust grows rapidly with the onset of transparency in the ultraviolet (UV) and as the optical depth decreases with the expansion-driven dilution following the explosion. The challenge is therefore to be able to explain the formation of large grains on rather short timescales. In comparison, in the WR 140 system the typical grain radii appear to be of the order of $30-40$nm \citep{2023ApJ...951...89L} and must also be formed on similarly short timescales. For grains forming in the physical conditions typical of nova and WR systems indicated in Table \ref{table_conds} charge disproportionation occurs around a neutrality condition of $\delta_{\pm} |Z_{\rm d}| = 2$ (i.e. $Z_{\rm d} = -1$ to $+1$) and as large as $\delta_{\pm} |Z_{\rm d}| = 8$ ($Z_{\rm d} = -1$ to $+7$), resulting in Coulomb enhancement factors of $\sim 3-5$ for the lower density case in Fig. \ref{fig_charge} (cobalt line). This is the most interesting of the cases studied because it represents a dust formation stage where an outflow is beginning to undergo ionisation by low energy UV photons ($E_{{h}\nu} \leqslant 6$eV) and the density is decreasing with the expansion of the explosion or outflow. Of note is that the collisions of nucleating clusters ($a \simeq 0.3$nm, $Z_{\rm d}=-1$), consisting of only a few carbon atoms, with small grains ($a \simeq 3$nm, $Z_{\rm d}=+1$, cobalt line Fig. \ref{fig_charge}) will be enhanced by a factor of $\sim 5$ with respect to the geometrical cross-section.  For particles somewhat larger ($a \simeq 1$nm, $Z_{\rm d}=-1$) in collision with grains further along in the condensation sequence ($a \simeq 40$nm, $Z_{\rm d} =+7$, blue line Fig. \ref{fig_charge}) the Coulomb collisional enhancement factor is $\sim 3$. While these factors are modest within the scheme of grain growth, they do indicate an enhancement in the dust nucleation and formation rates in nova and WR circumstellar environments.  

Fig. \ref{fig_coll} shows that 
%\begin{color}{blue}
%the most frequent collision partner with the smallest particles ($a \lesssim 0.3$nm) are C atoms, closely followed by Ca$^+$ ions aided by the large Coulomb factor ($\gtrsim 100$) which counteracts the low Ca atom abundance ($10^{-5}$). This figure also shows that O atom collisions, assuming $X$(O) $= 8 \times 10^{-4}$, with the smallest clusters ($0.2-0.3$nm, with $\simeq 3-10$ carbon atoms) are almost as frequent as those of the Ca$^+$ ions. Thus, the carbonaceous dust-nucleating clusters interact with a Ca$^+$ ions and C and O atoms over timescales of the order of days to a few weeks at the onset of ionisation (for the parameters given in Table \ref{table_conds}). 
%\end{color} 
%Further these 
small clusters ($Z=-1$) interact with, and could be accreted onto larger grains ($a >2$nm, $Z=+1$), on timescales of the order of a tens of collisions per day due to the Coulomb factor enhancement. Thus, it would appear that as soon as the small, anionic clusters form they will likely be swept up by somewhat larger, cationic particles. A result that would seem to predict a granular structure for the dust particles that form and grow in nov\ae, consistent with the nova stardust analysis of \cite{2019NatAs...3..626H}. 

%\begin{color}{blue}
%From Fig. \ref{fig_charge} and Table \ref{table_conds} it can be seen that the disproportionation radius decreases with decreasing density in the dust nucleation cases. The metals Ca, K and Na will ionise as the gas becomes transparent in the UV ($E_{{\rm h}\nu} < 7$eV) and this will in turn be associated with grain charging and the appearance of disproportionation, triggering dust nucleation and cluster accretion onto larger particles. However, as the density in the outflowing gas decreases the disproportionation radius will shift to ever smaller radii until reaching a point where it can no longer be sustained and all grains will be positively charged. Within this scenario there would therefore seem to be a somewhat narrow window, a sweet spot, for rapid dust nucleation and growth, with timescales seemingly of the order of days to weeks. 

%In nov\ae\ the rapid growth of dust is associated with the onset of UV transparency, in the declining optical light curve, and is clearly triggered immediately before the optical decline towards the minimum due to the opacity of newly formed dust, the so-called dust dip which correlates with the excess in the infra-red emission arising from that same dust. After this epoch of dust formation the opacity reaches a maximum before decreasing due to dilution in the expanding outflow  \citep{Shore:2004hj,Shore:2018bi}. 
%\end{color} 

Given the decrease in their optical light curves, with minima  at about 100 days after outburst, nova would seem to rapidly form dust over a timescale of 60 to 150 days \citep{Shore:2004hj,Shore:2018bi}. Thus, for any viable dust nucleation and growth process it must be capable of forming dust on these sorts of timescales. To fully investigate the possible role of disproportionation in dust nucleation will require detailed physical modelling, with a full size-dependent dust charge calculation, something that is beyond the scope of this simple exploratory study.

%-------------------------------------------------------------------------------------------------------------------------------- 
\subsection{Dust growth in the ISM}
\label{sect_ISM}
%-------------------------------------------------------------------------------------------------------------------------------- 

The results for dust collisions in the CNM and in the outer regions of dense clouds as a function of $A_{\rm V}$ are shown in Fig. \ref{fig_charge}. In the CNM all grains, except the very smallest clusters ($a = 0.2$nm), are positively charged and the Coulomb factors are repulsive,  as indicated by the CNM charge distribution shown by the thin red line in Fig. \ref{fig_charge}. We note that the smallest clusters, containing only few carbon atoms, will have very short lifetimes against photo-dissociation in the CNM \citep{Jones.Ysard.2025} where the minimum carbon nanoparticle radius predicted by THEMIS is 0.4nm. 

The results presented in Section \ref{sect_ISM_growth} suggest that size-dependent charge disproportionation could help to explain the rapid disappearance of carbonaceous nanoparticles in the outer layers of molecular clouds in regions with densities of the order of  a few 1000\,cm$^{-3}$ and $A_{\rm V} \simeq 1$.

Modelling studies \citep{2003A&A...398..551S,2012A&A...548A..61K,2015A&A...579A..15K} indicate that the observed disappearance of the stochastically-heated nano-grains in dense regions \citep{1991ApJ...372..185L,1996A&A...309..245A,1994ApJ...423L..59A,1998A&A...333..709L,2003A&A...399.1073K,2003A&A...398..551S,2012A&A...548A..61K} can be explained by nano-grains coagulating onto larger grains. Further, \cite{2020A&A...641A..39S} needed to assume perfect sticking during collisions to remove all the small grains in order to explain the observations of dust in clouds. This was also an underlying assumption in the work of \cite{2013A&A...559A.133Y} and would seem to be coherent with the conclusions arrived at by \cite{2016A&A...588A..43J} and \cite{2016A&A...588A..44Y}. The latter in their coreshine and cloudshine modelling needed to assume that all small grains have already disappeared from within a cloud at an A$_{\rm V}$ of unity. Following on from these results, we speculate that they may help to explain why the formation of rotationally supported protostellar discs, in simulations, is only possible if grains smaller than 1nm are removed from the dust size distribution from the very beginning, as revealed by \cite{2016MNRAS.460.2050Z} and further discussed by \cite{10.3389/fspas.2022.949223}.

\end{appendix}

\end{document}